\documentclass{ifacconf}

\usepackage{graphicx}      
\usepackage{natbib}        
\usepackage{amsmath,amssymb,amsfonts}
\usepackage{adjustbox}
\usepackage{array}
\usepackage{booktabs}
\usepackage{rotating}
\usepackage{breqn}
\usepackage{xcolor,colortbl}
\definecolor{Gray}{gray}{0.92}

\newcolumntype{P}[2]{%
  >{\begin{turn}{#1}\begin{minipage}{#2}\small\raggedright\hspace{0pt}}l%
  <{\end{minipage}\end{turn}}%
}
\newcolumntype{R}[2]{%
>{\columncolor{Gray}}
  >{\begin{turn}{#1}\begin{minipage}{#2}\small\raggedright\hspace{0pt}}l%
  <{\end{minipage}\end{turn}}%
}
\newcolumntype{a}{>{\columncolor{Gray}}c}

\begin{document}
\begin{frontmatter}


\title{A multifactorial evaluation framework for gene regulatory network reconstruction\thanksref{footnoteinfo}}

\thanks[footnoteinfo]{AA and JM are supported by University of Luxembourg internal researh projects PPPD (both) and OptBioSys (AA).}

\author[First]{Laurent Mombaerts} 
\author[First]{Atte Aalto} 
\author[First]{Johan Markdahl}
\author[First]{Jorge Gon\c{c}alves}

\address[First]{Luxembourg Centre for Systems Biomedicine, University of Luxembourg, 
   6 avenue du Swing, 4367 Belvaux, Luxembourg (e-mail: laurent.mombaerts@uni.lu, atte.aalto@uni.lu, johan.markdahl@uni.lu, jorge.goncalves@uni.lu).}

\begin{abstract}                
In the past years, many computational methods have been developed to infer the structure of gene regulatory networks from time-series data. However, the applicability and accuracy presumptions of such algorithms remain unclear due to experimental heterogeneity. This paper assesses the performance of recent and successful network inference strategies under a novel, multifactorial evaluation framework in order to highlight pragmatic tradeoffs in experimental design. The effects of data quantity and systems perturbations are addressed, thereby formulating guidelines for efficient resource management.\\

Realistic data were generated from six widely used benchmark models of rhythmic and non-rhythmic gene regulatory systems with random perturbations mimicking the effect of gene knock-out or chemical treatments. Then, time-series data of increasing lengths were provided to five state-of-the-art network inference algorithms representing distinctive mathematical paradigms. The performances of such network reconstruction methodologies are uncovered under various experimental conditions. We report that the algorithms do not benefit equally from data increments. Furthermore, for rhythmic systems, it is more profitable for network inference strategies to be run on long time-series rather than short time-series with multiple perturbations. By contrast, for the non-rhythmic systems, increasing the number of perturbation experiments yielded better results than increasing the sampling frequency. We expect that future benchmark and algorithm design would integrate such multifactorial considerations to promote their widespread and conscientious usage.
\end{abstract}


\end{frontmatter}

\section{Introduction}

Genes in living organisms do not function in isolation, but may activate or suppress the activity of other genes, forming intricate networks of regulatory interactions. The investigation of the struture of gene regulatory networks (GRN), which consists in unveiling the set of biochemical interactions that control gene expression, is a fundamental step in the understanding of disease mechanisms and the development of future drugs and therapies. Recovering the topology of such networks from gene expression profiles remains, however, a crucial challenge in systems biology due to their vast complexity, intrinsic stochasticity, and limited observability.


In the last few years, many computational methods based on a variety of mathematical paradigms have been developed to reconstruct GRNs from time-series data, and their performances assessed on relevant mathematical models \citep{ARNI, Aderhold_mechanistic, Marbach2012}.
However, experimentalists are still faced with difficult questions: Which method to use with an available dataset? On the other hand, with fixed amount of resources, what kind of experiments to carry out to ensure optimal use of resources? How much can be gained by investing into a few more measurements?




The purpose of this study is to provide guidelines for conscientious management of biological resources by unveiling the performances of state-of-the-art network inference strategies under various experimental designs. To this end, the effects of data quantity and multi-experiment availability are assessed simultaneously on the accuracy of the topological reconstruction.

Navigating experimental tradeoffs for GRN inference is not an entirely new concern, although literature on the topic is surprisingly scarce \citep{HAQUE201996}. Such a multifactorial approach has never been undertaken for systematic evaluation of network reconstruction algorithms. \cite{Sefer} studied the tradeoffs between dense and replicate sampling strategies. Their results showed that, under reasonable noise assumptions, gene expression profiles reconstructed from dense sampling are more accurate than those resulting from replicate sampling. \cite{geier2007reconstructing} showed that at equivalent data size, short-time, gene knock-out experiments contain more information about the GRN structure than single experiment, longer recordings of non-rhythmic systems. The GRN inference algorithms used in their study, however, are no longer state-of-the-art. \cite{Markdahl_CDC} studied the  cell cycle of Saccharomyces cerevisiae as a case-study to analyze the effect of temporal resolution on the quality of the inferred network. The performance as a function of time-series length resulting from a LASSO methodology \citep{lasso} resembled a sigmoid shape with a plateauing effect at the end. \cite{Mombaerts_FOSBE} used a model of the circadian network of Arabidopsis thaliana, a rhythmic system with period of about 24 hours, to show that sampling from transient dynamics provides richer information for the identification of the underlying GRN topology. \cite{Muldoon} identified previously unrecognized factors that affect inference outcomes, such as stimulus-specific experimental design and network motifs in the vicinity of a stimulus. Following the DREAM3 competition, \cite{GRN_review} investigated strengths and weaknesses of algorithms in recognizing types of motifs that appear in gene networks. Finally, it has also been shown that, for mutual-information based techniques, the accuracy reaches a saturation point after a specific data size \citep{Chaitankar2010}. Algorithms based on correlation or mutual information are, however, excluded from this research as they cannot detect causality between genes \citep{Marbach2012}. 

Realistic time-series datasets were generated from one rhythmic and five non-rhythmic models of gene regulatory networks that have been widely used as benchmarks in the literature \citep{Aderhold_mechanistic, DREAM_models}. External interventions, i.e., gene deletion (knock-out) and chemical treatments, are explicitly simulated to provide a comprehensive picture of the performance of each algorithm under a range of experimental conditions. Increasingly rich multi-experiment time series datasets are fed to five network inference algorithms. Performance is assessed using standard techniques for classification algorithms, studying the area under the receiver operating characteristics (ROC) and the precision-recall (PR) curves. We expect such pragmatic considerations to contribute to the development of conscientious experimental designs and appropriate mathematical benchmarks. 

\section{Methods}

\subsection{Data generation}

The use of in silico networks is preferable to random graphs, as they account for realistic structural properties of biological networks \citep{Schaffter2011}. The oscillating gene regulatory system used as a benchmark in this paper is a model of circadian regulation of Arabidopsis thaliana, hereafter referred to as Millar 10 (Figure 1A) \citep{Millar10}. Conceptually, it is composed of 2 interconnected feedback loops and an input pathway that incorporates external light cues in order to synchronize the plant to the surrounding light conditions. When subjected to another external regime, e.g., constant light or constant darkness, the system displays transient dynamics and reaches a new limit cycle. It is expected that measurements of the transient period contain more information than measurements of a limit cycle. This hypothesis was confirmed by \cite{Mombaerts_FOSBE}. To complement these results,
the performances of the network inference strategies are first analyzed under such transient dynamics. For this purpose, the model has been simulated for 240 hours in light/dark cycles to remove initial system transients and then switched to constant light regime. Time windows of 48 hours are then extracted from the light/dark limit cycle, at the transition to constant light and 48 hours after transition to constant light for further comparison, emulating the computational framework of \cite{Mombaerts_FOSBE}. In the following analysis, time windows starting from the transition to constant light and up to 3 days of observations (24-36-48-60-72 hours) are considered.

\begin{figure}
\centering
\includegraphics[width=8.8cm]{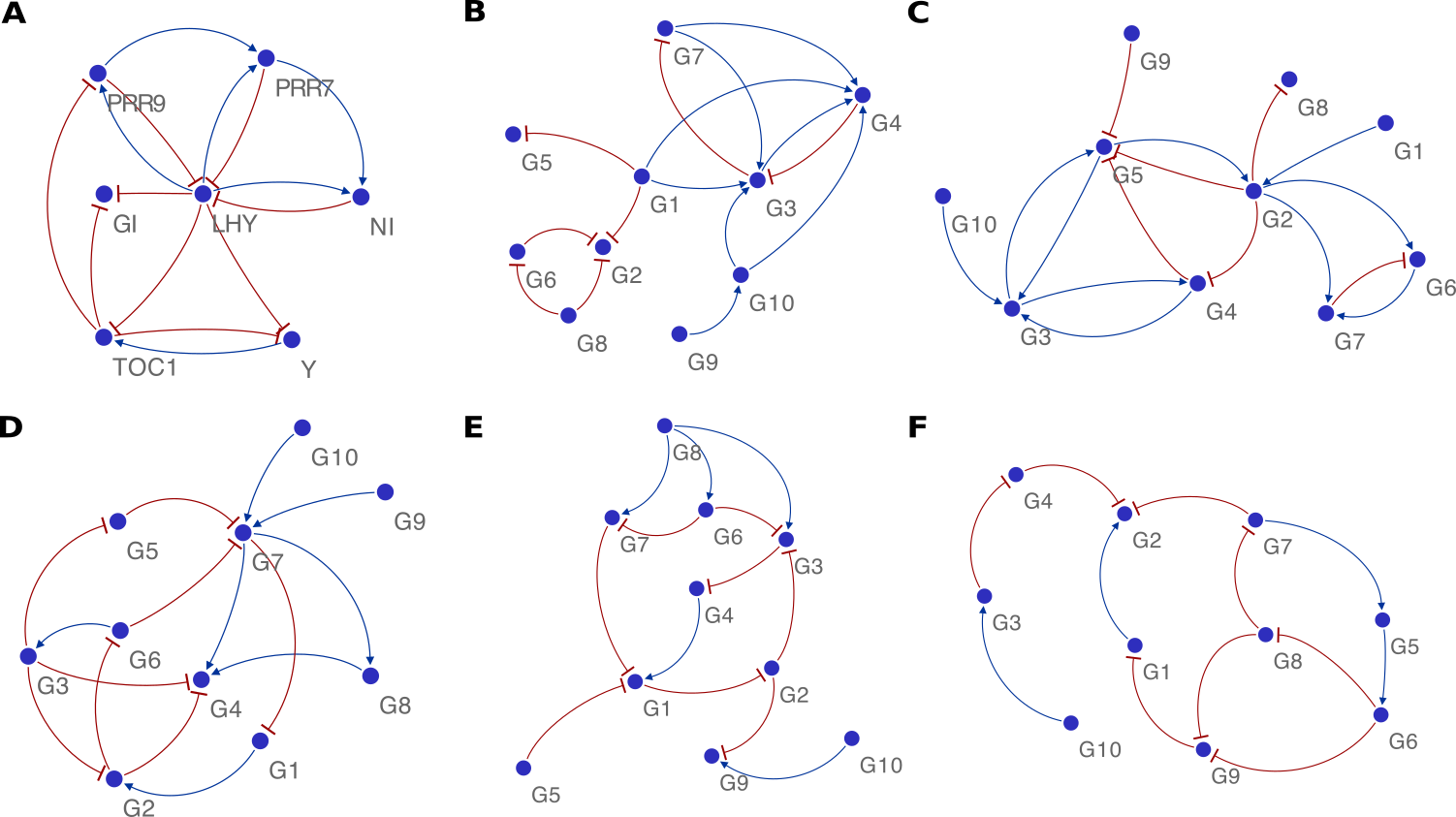}
\caption{Gene regulatory networks used as benchmarks. Blue pointed arrows and red blunt arrows represent activation and inhibition reactions respectively. \textbf{A.} Millar 10 Model (Rhythmic) \textbf{B.--F.} DREAM4  models 1--5}
\label{fig:nets}
\end{figure}

The simulations are based on Langevin equations (stochastic differential equations) to model the intrinsic stochasticity in the dynamics of gene regulatory networks (accounting for molecular noise in both transcription and translation processes) \citep{Langevin}. The intrinsic noise is expected to have a significant impact on the behavior of the system \citep{Stochastic_Millar}. Then, data are downsampled to resemble realistic experimental design (every 4 hours in the case of circadian experiments). Finally, the protein concentrations are not made available in the provided datasets (only mRNA concentrations levels). 

The Millar 10 model has been simulated to reproduce gene knock-out experiments. Knock-out experiments are very informative, more than knock-down experiments, as they provide network response to individual and large perturbations (genes are deleted) \citep{Marbach2012}. Knock-out experiments were simulated as in \cite{Aderhold2014}, by replacing the transcription rates of the targeted genes by random noise drawn from a truncated normal distribution to ensure non-negativity of the concentrations. Genes that have been knocked out are, thefore, not influenced by their structural regulators anymore. The datasets in the numerical experiments consist of a wildtype (WT) time series, and up to three randomly chosen knock-out time series at a time. This selection has been randomized 6 times to account for the uneven informative potential of different genes in the network. Furthermore, such simulations being stochastic, each experiment was replicated 10 times to provide a representative view of the performance of each computational method. In total, $950 = 10 \cdot 5 +  10 \cdot 6 \cdot 5 \cdot 3$ simulations were performed.

The studied non-rhythmic models (Figure 1 B--F) originate from the DREAM4 in silico network inference challenge \citep{DREAM_data,GRN_review,DREAM_models}. New time series data were generated by the GeneNetWeaver software \citep{Schaffter2011} which simulates the system's response to a perturbation of about a third of its nodes, followed by a relaxation back to steady state after half of the recording, when the perturbation was removed. The data characteristics are as in the challenge, with the exception that the perturbation targets were randomized, whereas in the challenge data they were preferentially carried out to cover the whole network \citep{Marbach2012}. 
Such data simulation offers a realistic representation of the undetermined effects of a chemical treatment on a system at rest (steady-state). In essence, the modeling specificities are similar to those of the Millar 10 model in the sense that both transcription and translation are modeled, only the mRNA concentration levels are made available and the simulations are based on Langevin stochastic equations. The resulting data are then resampled using 3 different sampling rates (11-21-41 datapoints) to further assess the effect of changes in experimental design. Here, 10 chemical perturbations were simulated for each of the 5 available networks and replicated 3 times. Then, increasing amount of perturbation time series (up to 4 at a time)  are randomly selected from those 10 generated perturbations, and provided to the inference algorithms. This random selection is performed 5 times. 
The perturbation targets are not known to the methods. In total, $900 = 3\cdot5\cdot3\cdot4\cdot5$ simulations were performed.

\subsection{Network inference methods}


Formally, the reconstruction of the structure of gene regulatory networks corresponds to the identification of the regulators $\mbox{\boldmath$\pi$}_i$ for all given genes $i \in \{1,..,n\}$ (where $n$ is the amount of genes in the system), so that
\begin{equation}
\frac{dy_i(t)}{dt} = f(\mbox{\boldmath$\pi$}_{i}(t)) - \alpha_i y_i(t)
\end{equation}
where $y_i(t)$ is the mRNA concentration of gene $i$ at time $t$ and the term $\alpha_i y_i(t)$ corresponds to the degradation rate of $y_i(t)$. The nonlinear function $f$ represents the influence of the transcription factors of the parents on the target genes. In practice, however, the time course data of the regulators $\mbox{\boldmath$\pi$}_i$ consist of gene expression levels as the protein levels are typically not available to us. Most methods then consider a simplified model (1) where gene expression values of the parent genes are used as proxies to transcription factors, and then $\mbox{\boldmath$\pi$}_i(t) \subset \{ y_1(t),...,y_n(t)\}$. The concatenation of the structural regulators $\mbox{\boldmath$\pi$}_i$ of every gene, then, forms the network underlying gene regulation, which is mostly sparse. The problem is typically undetermined, since the number of available time points is small compared to the number of possible solutions and since not all species are observable. Moreover, regulatory interactions can either be additive or multiplicative, while proteins might undergo post-translational modifications, adding another layer of complexity. The difficulty for the development of inference algorithms is to decide upon the complexity of the strategy while being aware of the inherent overfitting danger. We aim to highlight the advantages and optimal applicability ranges for current state-of-the-art methodologies by estimating their performances on various experimental scenarios. While the results presented are by no means exhaustive in terms of strategies or experimental design scenarios, they support the efficient use of biological resources.

The network inference methods included in the comparison represent the function $f$ at various levels of complexity. They are All-to-all (ATA) \citep{DyDE}, Gaussian Process Dynamical Models (GPDM) \citep{GPDM_GRN}, dynamical GEne Network Inference with Ensemble of trees (dynGENIE3) \citep{dynGENIE3}, Algorithm for
Revealing Network Interactions (ARNI) \citep{ARNI} and Improved Chemical Model Averaging (iCheMA) \citep{Aderhold_mechanistic}. For details, we refer to publications presenting the methods. 

The All-to-all method is a parametric method based on fitting a linear model (of order one, by default, but higher order models are also possible) i.e., $f$ is linear in (1), between each pair of genes, one pair at a time. A fitness score is computed for each pair, which is regarded as a confidence level on the existence of the corresponding regulation. Here, the methodology presented by \cite{DyDE} has been extended to deal with multi-experiment datasets. The datasets are merged together so that the dynamics to be identified are identical through all experimental conditions, assuming that the signal-to-noise ratios are similar in different experiments. The fitness score over multiple experiments is
\[
\scriptstyle
\mathrm{fit} \, = \,  1 - \frac{\sqrt{\sum_{t \in T^1} (y_1(t) - \hat{y}_1(t|\theta))^2} + ... + \sqrt{\sum_{t \in T^n} (y_n(t) - \hat{y}_n(t|\theta))^2}}{\sqrt{\sum_{t \in T^1} (y_1(t) - \bar{y}_1(t))^2} + ... + \sqrt{\sum_{t \in T^n} (y_n(t) - \bar{y}_n(t))^2}}
\]
where $n$ represents the amount of experimental conditions, $T$ the sampling times of experiments, $y$ the gene expression level, $\hat{y}$ the modeled gene expression, and $\bar{y}$ is the average value of the gene expression level.

GPDM, a non-parametric method, models gene expression as a nonlinear stochastic differential equation
\[
\begin{cases}
y_j = x(t_j) + v_j \\
dx = g(x,\theta) dt + du
\end{cases}
\]
where the dynamics function $g$ is modeled as a Gaussian process with some hyperparameters $\theta$, $v_j$'s correspond to measurement errors, and $u$ is a Brownian motion. This defines gene expression as a stochastic process whose realizations can be sampled using a Markov Chain Monte-Carlo (MCMC) strategy. Network inference is based on estimating the hyperparameters of the covariance function of the GP. Multi-experiments are taken into account by assuming that all time series are produced by the same dynamics function $g$. Independent samplers are then constructed for trajectories $x$ corresponding to different experiments. Performance of GPDM was recently compared to the best performers of the DREAM4 challenge and consistently shown superior in dealing with short time series data \citep{GPDM_GRN}.

ARNI is a recently developed non-parametric method used for the estimation of network topologies that performed well in network inference from a large collection of short time series. The derivatives are estimated explicitly through a difference approximation and the relationships between nodes in the network are estimated by solving a nonlinear regression problem, with a user-selected library of nonlinear basis functions, using a greedy approach \citep{ARNI}. In the experiments, we used polynomial basis functions of degree at most 3. 

The semi-parametric method dynGENIE3 is an adaptation of the GENIE3 method for time series data. GENIE3 was the best performer in the DREAM4 Multifactorial Network Inference challenge and the DREAM5 Network Inference Challenge. The transcription function $f$ in (1) is represented by an ensemble of regression trees which is estimated from the gene expression data and their derivatives, estimated using a difference approximation \citep{Huynh-Thu2018}. 

Alternatively, iCheMA estimates derivatives from the data by fitting a smooth Gaussian process to the time-series. Then, gene expression profiles are modeled using the Michaelis-Menten formula for mass-action kinetics:
\[
\frac{dy_i(t)}{dt} = \sum_{u \in \pi_i} v_{u,i} \frac{I_{u,i}y_{u}(t) + (1 - I_{u,i}) k_{u,i}}{y_{u}(t) + k_{u,i}}- \alpha_i y_i(t)
\]
where $k_{u,i}$ corresponds to the Michaelis-Menten parameters and $I_{u,i}$ indicates whether the regulation is an activation ($I_{u,i} = 1$) or an inhibition ($I_{u,i} = 0$). Network inference is then based on estimating the parameters using an MCMC approach. iCheMA goes exhaustively through all possible combinations of regulators (typically, up to 3 at a time), which makes it a computationally heavy algorithm that does not scale easily to large systems.

\begin{table}[b]
\caption{Properties of the different methods.}
 \centering
\footnotesize
  \begin{tabular}{lacacaca}
 Method     & \multicolumn{1}{R{90}{1.9cm}}{Nonlinear dynamics} & \multicolumn{1}{P{90}{1.9cm}}{Continuous-time} & \multicolumn{1}{R{90}{1.9cm}}{Combinatorial effects} & \multicolumn{1}{P{90}{1.9cm}}{\raisebox{-3.7mm}{Hidden nodes}} & \multicolumn{1}{R{90}{1.9cm}}{Computation time (s)} & \multicolumn{1}{P{90}{1.9cm}}{Performance \tiny{Millar/DREAM}} \\
   \hline 
    All-to-all & & ${\small \checkmark}$ & & $({\small\checkmark})^1$ & 49.4 & 2/5 \\
   GPDM & ${\small \checkmark}$ & ${\small \checkmark}$ & ${\small \checkmark}$ & & 333.4 & 1/1 \\
   dynGENIE3 & ${\small \checkmark}$ & & ${\small \checkmark}$ & & 0.7 & 4/2 \\
   ARNI & ${\small \checkmark}$ & & $({\small \checkmark})^2$ & & 1.0 & 3/3 \\
   iCheMA & ${\small \checkmark}$ & & $({\small \checkmark})^2$ & & 1999 & 5/4 \\
    \hline
    \multicolumn{7}{l}{$^1$If higher-order dynamics are used.} \\
\multicolumn{7}{l}{$^2$Discussed in the article, but not in the implementation.}
     \end{tabular}
       \label{tab:methods}
       \end{table}

Table~\ref{tab:methods} summarizes the properties of the methods included in the comparison. A method is deemed a continuous-time method if it is based on continuous trajectory-fitting, or modeling from a continuous-time system. Methods estimating derivatives from the data and then solving input-output regression are deemed discrete-time methods. Combinatorial effects mean dynamics of the form $\dot y_i = f(y_j,y_k)$ where $f(y_j,y_k)$ cannot be represented as a sum $f(y_j,y_k)=g(y_j)+h(y_k)$. The table shows whether the methods explicitly take into account combinatorial effects. The computational time is based on 48 hours recordings (13 datapoints) of the Millar 10 model. Performance ranking is based on average AUPREC values.

\section{Results}

The performances of each algorithm are here assessed in terms of the resulting Area Under the Receiver Operating Characteristic (AUROC) and the Area Under the Precision-Recall (AUPREC). Precision-Recall curves allow for a more accurate picture of algorithms performances for sparse GRNs and is commonly used for the comparison of inference algorithms. Auto-regulatory interactions are not considered.

Decomposing the time-series resulting from the rhythmic model into synchronized-desynchronized states showed that, on average, the accuracy of the network reconstruction is improved by considering transient dynamics (Figure~2). While GPDM, ATA, and dynGENIE3 benefit ---to a varying degree--- from the transition to the desynchronized state, change in performance of iCheMA was not statistically significant and ARNI's performance was slightly impaired. It should be noted that a significant increase in accuracy is observed for the strategies that do not explicitly estimate derivatives.

Figures 3 and 4 display the performance of each algorithm resulting respectively from the simulations with data from the Millar 10 model and the steady-state systems under several combinations of data types. 

\begin{figure}
\includegraphics[width=9cm]{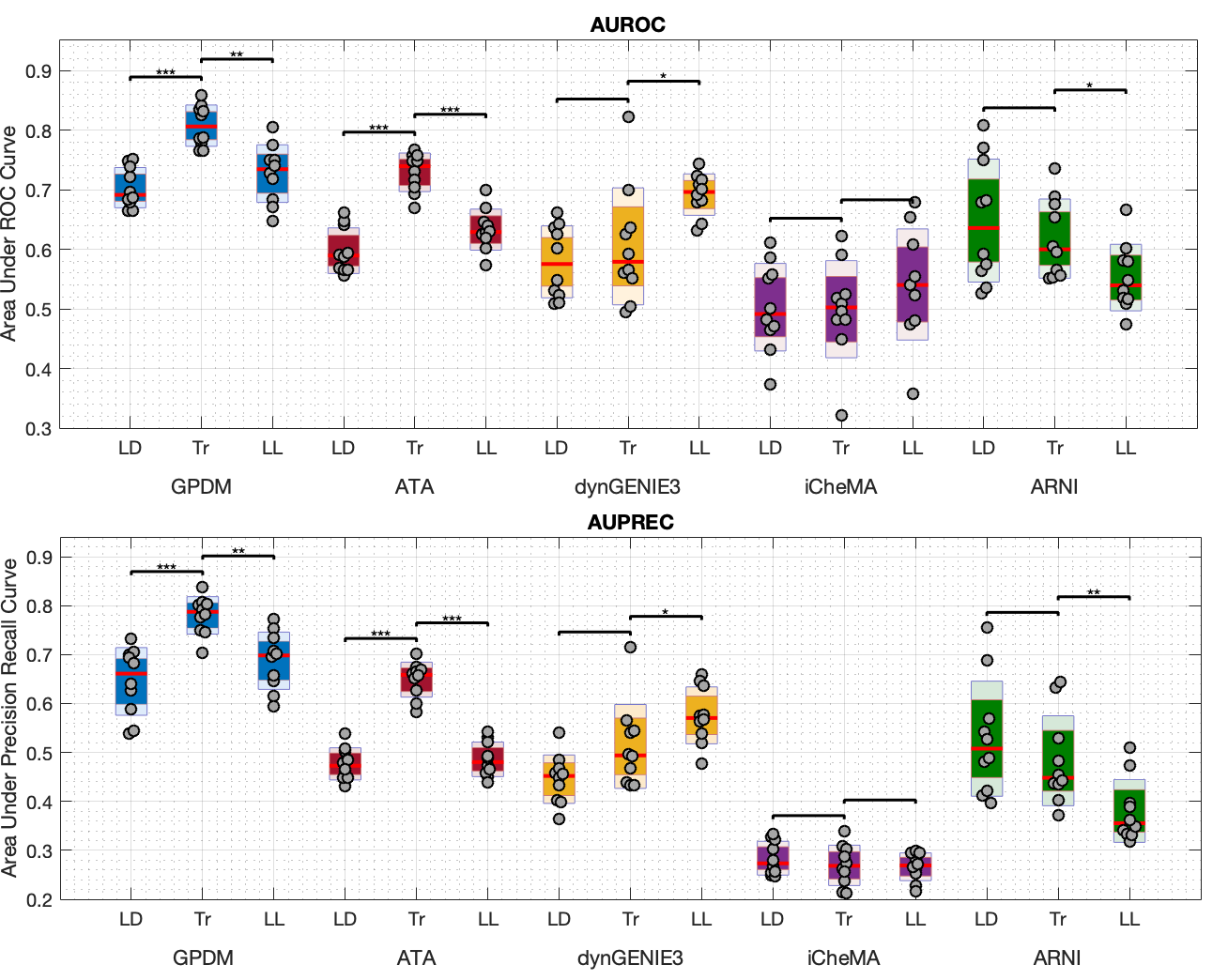}
\caption{Evaluation of the effects of dynamical transients on the rhythmic model (Millar 10). Statistical significance between predictions from transient dynamics and other conditions are indicated by star symbols. They are computed with the Mann-Whitney U-test.}
\label{fig:nets}
\end{figure}

On one hand, these graphs show that GPDM outperforms the other approaches for almost every system and experimental configuration considered. It is outperformed by ATA in only one case with 24h wildype only data from the Millar 10 model. This illustrates the importance of various experimental scenarios in benchmarking network inference strategies and motivates the work undertaken in this paper. Interestingly, the simple pairwise low order linear modeling (ATA) seems to outperform dynGENIE3, iCheMA and ARNI in terms of AUPREC for every observation length and system perturbation considered in the rhythmic model. Only the AUROC values of the non-parametric approach ARNI exceed those of ATA for the 3 mutations case, starting from 36 hours of observation. 

It is further interesting to notice that not all algorithms benefit equally from data increments. The gain of accuracy resulting from increasing the amount of data in the rhythmic model is only mild for the linear modeling strategy and iCheMA while it is significant for GPDM, dynGENIE3, and ARNI. In this regard, in average, GPDM benefits from the largest increase in accuracy whereas dynGENIE3 and ARNI compete at a slightly lower level for the experimental conditions presented. A saturation effect, however, can be observed at AUPREC values of around 0.8 for GPDM, 0.63 for the ATA, and 0.58 for ARNI.

The analysis of the DREAM competition models delivers a different view on network reconstruction as not all nodes are stimulated in a given system perturbation. For those networks, the benefit of additional system perturbations is considerable as they allow investigation of novel, previously unstimulated segments of the network. In the experimental design cases presented, none of the algorithms seemed to approach a saturation point for the data combinations considered. While the GPDM succeeds in providing the best accuracy for the DREAM networks as well, dynGENIE3 ranks second, ARNI third, iCheMA fourth and ATA last. The reason why the linear modeling strategy is surpassing dynGENIE3 and ARNI for the 1 perturbation only case and does not improve for additional datasets is likely related to the partially stimulated nature of the whole dynamical system and has yet to be investigated.

\begin{figure}
\includegraphics[width=9cm]{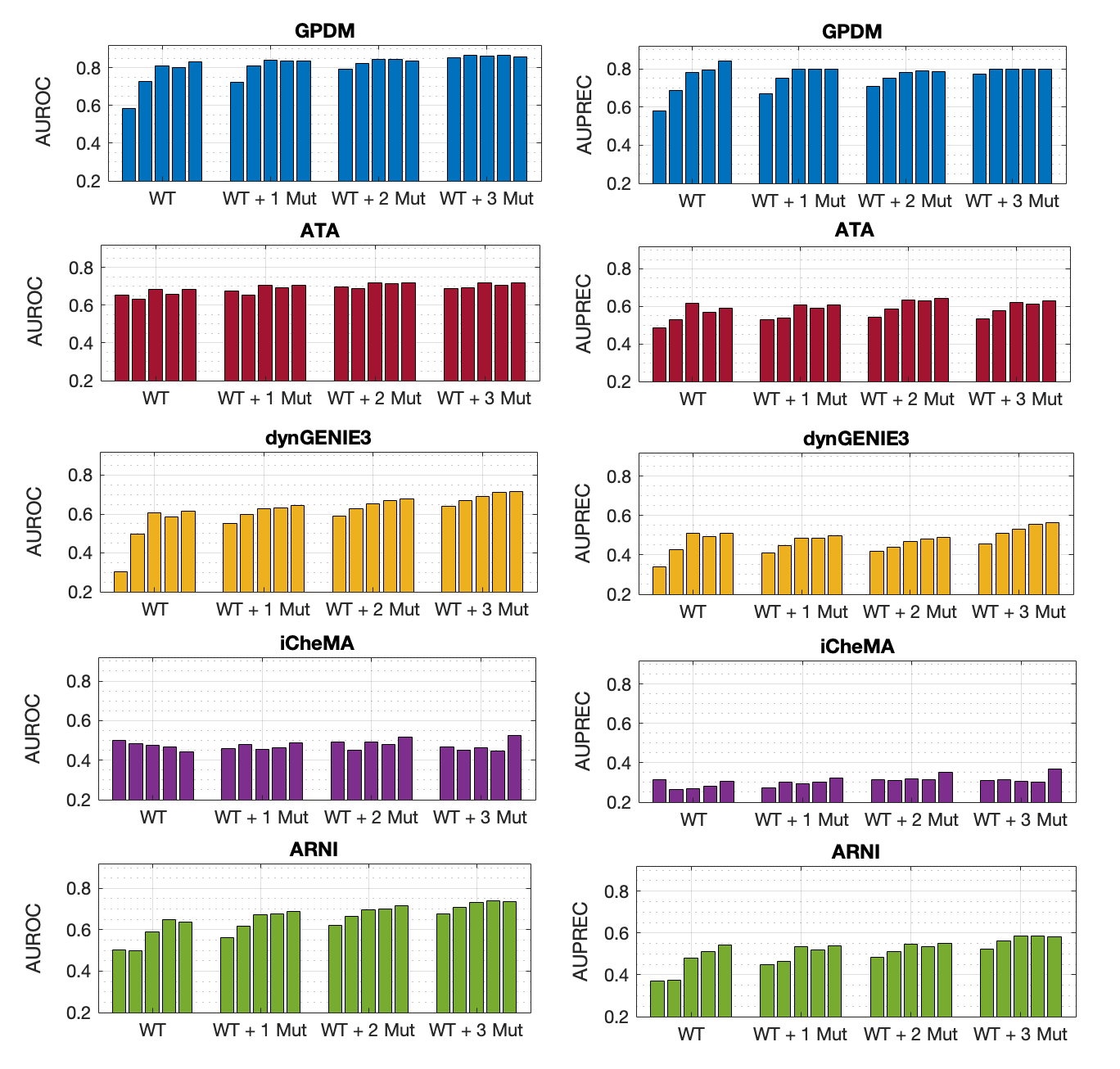}
\caption{Area Under the ROC curve and the Precision-Recall curve resulting from the inference of the Millar 10 model, for multiple combinations of observation lengths and system perturbations. The bars are grouped by the amount of additional recordings resulting from perturbations applied to the system (up to 3) and decomposed into data observation lengths of [24-36-48-60-72] hours (from left to right).}
\label{fig:nets}
\end{figure}

On the other hand, Figures 3 and 4 allow for proper visualization of experimental tradeoffs. Doubling the amount of datapoints by performing another experiment does not provide the same level of information than doubling the amount of datapoints in a given experimental setup. Table~2 summarizes the amount of datapoints in each of the presented experimental setups. While the cost of each datapoint might not be equivalent whether it originates from a novel system perturbation or from longer recording, these tables provide insight on how to choose an appropriate experimental scenario regarding the performances of each of the algorithms presented in Figures~3 and~4. For instance, sequencing a gene in WT every 4 hours during 72 hours requires a similar amount of datapoints as the WT with 2 mutations for 24 hours or the WT with 1 mutation for 36 hours. In this case, the experimental design that provides the best results would be a single recording of 72 hours resulting in an AUPREC of 0.84 using GPDM, compared to 0.75 or 0.7.
Regardless of the algorithm and assuming an equivalent cost per datapoint, it can be observed that, as a general rule of thumb and for top performing strategies, it is often preferable to observe the rhythmic system for a longer amount of time. By contrast, increasing the sampling frequency of the steady-state systems only resulted in a marginal improvement in the accuracy of network reconstruction. Surely, a lower bound on the sampling rate is required for a reliable construction of those systems but it was not reached in this analysis.

\begin{figure}
\includegraphics[width=9cm]{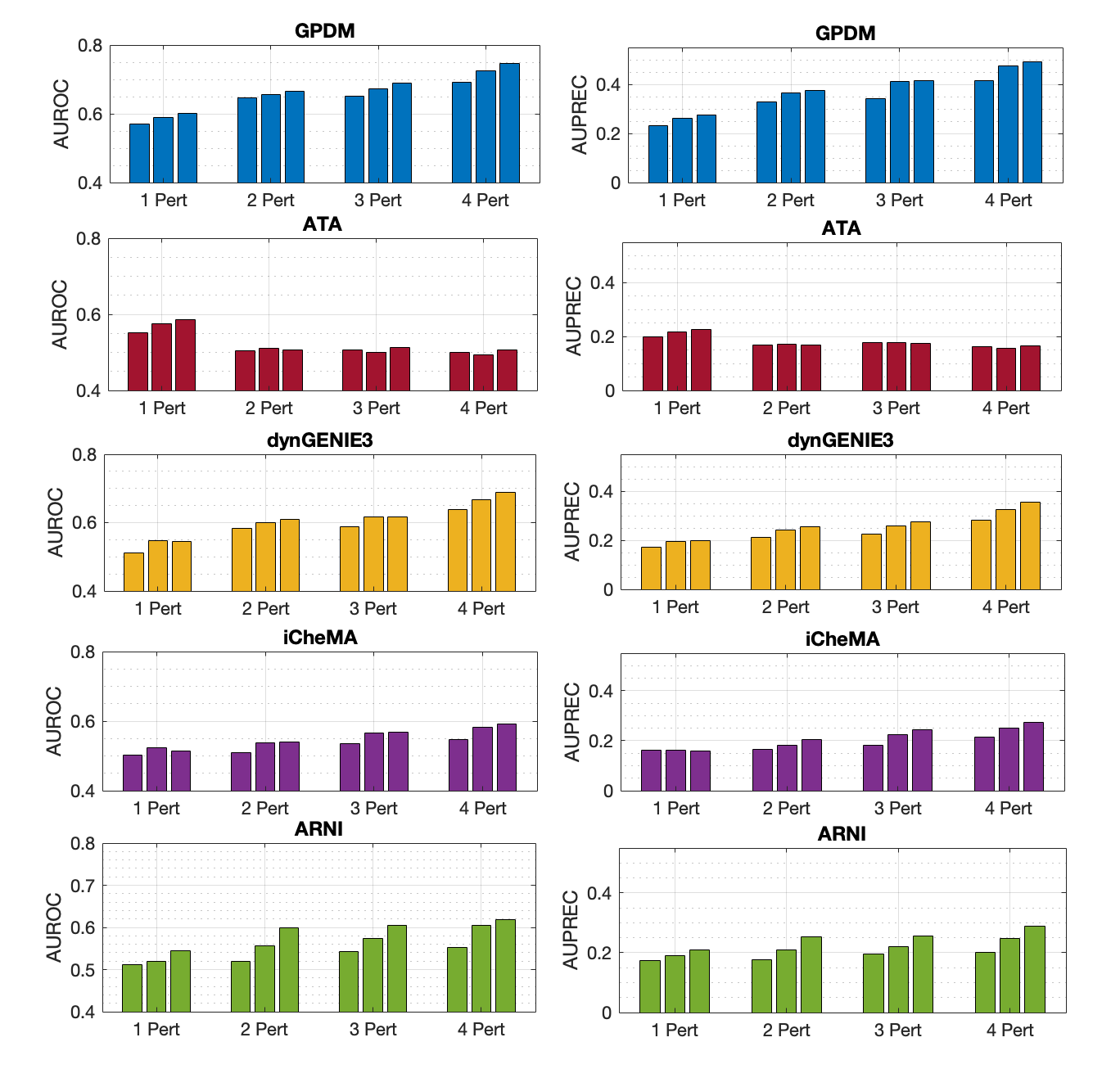}
\caption{Area Under the ROC curve and the Precision-Recall curve resulting from the inference of the DREAM models, for multiple combination of data sampling rates and systems perturbations quantity. The bars are grouped by the amount of systems perturbations (up to 4) and decomposed into data quantity of [11-21-41] datapoints (from left to right).}
\label{fig:nets}
\end{figure}

\begin{table}[t]
 \centering
 \caption{Left: Numbers of measurements in the Millar 10 experiments. Sampling rate is always 4h. Right: Numbers of measurements in the DREAM experiments. Window length is always 20.}
\scriptsize

\setlength{\tabcolsep}{5pt}
  \begin{tabular}{lccccc}
Window (h) \hspace{-2mm} & 24 & 36 & 48 & 60 & 72 \\ \hline
WT &       7 & 10 & 13 & 16 & 19 \\
WT + 1 Mut & 14 & 20 & 26 & 32 & 38 \\
WT + 2 Mut & 21 & 30 & 39 & 48 & 57 \\
WT + 3 Mut & 28 & 40 & 52 & 64 & 76 \\ \hline
\end{tabular}
\hspace{1.2mm}
\begin{tabular}{lccc}
$\Delta t$ & 2 & 1 & 0.5      \\ \hline
1 Pert &  11 & 21 & 41  \\
2 Pert & 22 & 42 & 82  \\
3 Pert & 33 & 63 & 123  \\
4 Pert & 44 & 84 & 164  \\ \hline
\end{tabular}
       \label{tab:data_Millar10}
       \end{table}

       

\section{Discussion}

Choosing between different experimental designs and network inference strategies depends on the research question and the resource constraints. For this purpose, performing a complete cycle study involving multiple inference strategies and specific benchmarks, such as in \citep{Madhamshettiwar2012}, is not uncommon. However, while such analysis provides a comprehensive idea of the relevance of the inferred network topology, it represents a significant investment of both time and money, and is sometimes not even possible.

In this paper, the general effects of data quantity and system perturbations on the accuracy of the GRN reconstruction were evaluated for one rhythmic model of gene regulation and five steady-state models. Our contribution is threefold. We showed the relevance of multifactorial benchmarks to assess the performances of network inference strategies, the importance of an appropriate choice of model complexity given data availability, and revealed pragmatic considerations for experimental designs. Depending on the cost of performing more experiments or increasing the amount of datapoints, one of those choices should be preferred. 

The algorithms considered here showed consistent performances across the 6 investigated networks. All network inference strategies did not, however, benefit equally from the increasing amount of data. Nevertheless, the fact that the parameter free, Gaussian process strategy GPDM has been consistently outperforming all strategies presented is noticeable and of further interest. In addition, by looking at the data expense and the resulting reconstruction accuracy, it should be further noted that GRN inference algorithms should improve the way various time series experiments originating from the same biological system are taken into account. 

\cite{Marbach2012} showed that, on average, a combination of network inference strategies leads to the best network reconstruction. As such, we noticed that the order in which the links were inferred by each algorithm, and experiment, was different. Further research, therefore, should learn the ranks, or confidence levels, of each link in the network reconstruction process and design a proper combination of the algorithms that optimizes their synergy, depending on the experimental conditions. 

Multifactorial studies such as the one presented here require a considerable amount of simulations. Some algorithms, such as those involving MCMC sampling, took several days to run on a 24 core workstation. As such, a complete analysis of the experimental design space is not possible but other decisional factors exist and require further inspection. Among those, \cite{Sefer} showed that denser sampling is preferable to additional replicates. Such strategy could be particularly profitable for transient data and to algorithms that explicitly estimate derivatives. \cite{Muldoon} used time series data originating from the DREAM 4 challenge to show that using only half of the perturbation data (without the recovery to steady-state) might be beneficial to some algorithms. Furthermore, some methods are able to incorporate information on external inputs, such as perturbations (with the targets still unknown), which increases the average performance. In addition, in practice, gene regulatory networks are often of bigger dimension which is not always accessible to the most computationally expensive algorithms. 

Finally, our study did not take into account prior knowledge of the system, which could potentially be iteratively integrated into each step of the network reconstruction. For example, a strategy such as the one presented by \cite{Ud-Dean2015} actively optimizes the precision of the predictions by proposing the next most informative knock out. In such case, the aforementioned results would likely understimate the resulting accuracy of the reconstruction. In fact, doubling the amount of data points by doubling the observation time or by performing an additional experiment not only provides different levels of information, but can reveal different parts of the network as well. Such strategy might be necessary to cope with the most isolated genes.

\bibliography{papers,bibs}
                                  
\end{document}